\def\listitem{\par \hangindent=50pt\hangafter=1
     $\ $\hbox to 20pt{\hfil $\bullet$ \hfil}}

\def\puncspace{\ifmmode\,\else{\ifcat.\C{\if.\C\else\if,\C\else\if?\C\else%
\if:\C\else\if;\C\else\if-\C\else\if)\C\else\if/\C\else\if]\C\else\if'\C%
\else\space\fi\fi\fi\fi\fi\fi\fi\fi\fi\fi}%
\else\if\empty\C\else\if\space\C\else\space\fi\fi\fi}\fi}
\def\SP{\let\\=\empty\futurelet\C\puncspace}

\def\h1{$h^{-1}$\SP}

\def\etal{{\it et al.\/}\ }

\def\eg{{\it e.g.\/}\rm,\ }

\parskip=\baselineskip
\documentstyle[12pt,aasms4,xtex,epsfig]{article}

\def\sp{\ }
\def\h{$h^{-1}$ Mpc\sp}

\def\etal{et al.\,}
\def\eg{e.g. \,}

\def\sig8 {$\sigma_8\>$} 
\def\xis {$\xi (s)\>$}

\def\void#1{{}}

\begin{document}

\title{Biasing in the Galaxy Distribution} 
\author{C. Benoist}
\affil{LAEC, CNRS, Observatoire de Paris-Meudon, 5 Pl. J. Janssen, 
92195 Meudon Cedex, France.}
 \affil{benoist@gin.obspm.fr}
\author{S. Maurogordato}
\affil{LAEC, CNRS, Observatoire de Paris-Meudon, 5 Pl. J. Janssen, 
92195 Meudon Cedex, France.}
 \affil{maurogor@gin.obspm.fr}
\author{L.N. da Costa\altaffilmark{1}}
 \affil{European Southern Observatory, Karl-Schwarzschild-Str.2, 
D-85748 Garching bei M\"unchen, Germany.}
\affil{ldacosta@eso.org} 
\author{A. Cappi}
 \affil {Osservatorio Astronomico di Bologna, via Zamboni 33, 
I-40126, Bologna, Italy.}
\affil{cappi@astbo3.bo.astro.it}
 \and
\author{R. Schaeffer}
 \affil{Service de Physique Th\'{e}orique, CEN-Saclay, F-91191, 
Gif-sur-Yvette Cedex, France.}
 \affil{schaeffer@amoco.saclay.cea.fr}
 
\altaffiltext{1}{and Departamento de Astronomia CNPq/Observat\'orio Nacional, 
rua General Jos\'{e} Cristino 77, Rio de Janeiro, R.J. 20921 Brazil.}

\begin{abstract} We investigate the variation of galaxy clustering
with luminosity using the recently completed SSRS2 sample. Clustering
measurements based on the two-point correlation function and the
variance of counts in cells reveal a strong dependence of clustering
on luminosity for galaxies brighter than $L_{*}$, while no significant
variation is detected for fainter galaxies. We derive a relative bias
versus magnitude relation which can be compared with theoretical
predictions.
Existing models of galaxy formation cannot adequately reproduce
the simultaneous steep rise of biasing at high luminosities and the
plateau at the low-luminosity end. Improved modeling of halo-galaxy
relation, and larger samples including low luminosities galaxies are
required to draw more definitive conclusions.

\end{abstract}


\section{Introduction}

Redshift surveys of galaxies have provided over the years a wealth of
information that has greatly contributed to our understanding of the
way galaxies are distributed in space. However, these observational
results on galaxy clustering are not easily used to constrain
theoretical models because of our poor understanding of the relation
between galaxy and mass distributions.

Although until recently the main emphasis of redshift surveys has been on 
the study of large-scale structure, several attempts have been made to
understand the dependence of galaxy clustering on their internal
properties such as luminosity, morphology, and surface brightness. An
understanding of the relative bias between different galaxy
populations is essential for constraining models of galaxy
formation. Most of these studies were based on the first generation of
redshift surveys like CfA1 (Huchra \etal 1983) and SSRS (Southern Sky
Redshift Survey) (da Costa \etal 1988) which were not suitable for this
kind of analysis because of the small volume surveyed and of the 
relative small number of galaxies in the samples. 

Not surprisingly the results have so far been rather controversial,
with some authors claiming evidence of luminosity segregation 
in the data (see for instance Hamilton 1988, Davis et al. 1988, B\"orner,
Mo and Zhou 1989, Dominguez-Tenreiro \& Martinez 1989, B\"orner,
Deng and Xia 1989, Maurogordato \& Lachi\`{e}ze-Rey 1991,
Maurogordato, Schaeffer and da Costa 1992) while others arguing against it
(as Phillipps \& Shanks 1987, Thuan et al. 1987, Alimi et al. 1988, 
Eder et al. 1989, Haynes \& Giovanelli 1989, Bingelli et al. 1990, 
Thuan et al. 1991, Hasegawa \& Umemura 1993).  While the dependence of
clustering on morphology is well established (Davis \& Geller 1976, 
Santiago \& da Costa 1990, B\"orner \& Mo 1990, Mo et al. 1992, Iovino 
et al. 1993, Dominguez-Tenreiro et al. 1994), the dependence on luminosity
is still a matter of debate.

Recently, a new generation of redshift surveys has been completed
calling for a new examination of the problem. Recent estimates of the
galaxy power spectrum from the CfA2 (Park \etal 1994), and of the
correlation function for Stromlo-APM (Loveday \etal 1995) redshift
surveys seem to show that some luminosity segregation does exist, but
a consensus has still not been reached over the range of magnitudes at
which the effect occurs.  In this paper we will analyze in detail the
dependence on luminosity of the correlation function and of the variance 
of counts in cells using the recently completed SSRS2. The
combination of dense sampling, large sky coverage and a large number of
galaxies allows us to define various subsamples to investigate
separately different effects that may have a bearing on our
conclusions. In section 2 we describe the sample and the corrections
applied to the data. In section 3 we discuss the different statistics
applied in our analysis, investigate the dependence of clustering on
luminosity and compare our results to those of previous work. In
section 4 we discuss the implications of our results on models of
galaxy formation. A brief summary of our main conclusions is presented
in section 5.

\section{The data}

In the analysis presented below we use the
 SSRS2 (da Costa et al. 1994), a magnitude-limited redshift survey centered
on the south galactic pole. It includes $\sim 3600$ galaxies and covers 
1.13 sr, delimitated by $-40^{o}<\delta<-2.5^{o}$ and $b<-40^{o}$, 
complete up to limiting magnitude $m_{B(0)} = 15.5$.
Radial velocities have been corrected to the Local Group rest frame 
(Yahil \etal 1977).

Since all galaxies in the SSRS2 have been assigned morphological
types following the ESO-Uppsala classification system ($T$), we have
adopted a morphological-dependent K-correction, $K(z,T)$, to compute
the absolute magnitude of a galaxy according to the expression

\begin{equation}
M = m - 25 -5 \log D_L - K(z,T)
\end{equation}
where $D_L$ is the luminosity distance for standard Friedman models:

\begin{equation}
D_L = \frac{c}{H_0} \frac{1}{q_0^2}
    [1 - q_0 + q_0 z + (q_0 -1) \sqrt{2 q_0 z +1}]
\end{equation}
 In the following analysis we have adopted $H_0 = 100$, $q_0 = 0.5$.
We have divided the sample in four morphological classes and applied
for each class the K-correction following  Efstathiou, Ellis
\& Peterson (1988):\\
\begin{center}
\begin{tabular}[t]{lcr}
E/SO
&$-5\le T\le 0$
   &$K=4.14z$\\

Sa, Sb, Sbc
        &$1\le T\le 4$
                &$K=2.90z$\\
Sc, Scd
        &$5\le T\le 6$
                &$K=2.25z$\\
Sdm
        &$7<T$
                &$K=1.59z$\\

\end{tabular}
\end{center}

Volume-limited samples were defined adopting a \lq\lq minimum
distance\rq\rq criterium by taking galaxies within a depth
corresponding to the smallest $D_{L}$ among those calculated for the
different morphological classes, which corresponds in our case to the depth 
of the early-type galaxies. Luminosity distances were then converted into
comoving distances.

\section{Analysis}

\subsection{Statistics}

We examine the clustering properties of
galaxies using the two-point correlation function $\xi(s)$ and the $J_3$ 
integral, computed by
counting the number of galaxy pairs in shells, and 
the variance $\sigma^2$ of galaxy counts in excess of Poisson derived 
from count probabilities in randomly placed shperes. 

The two-point correlation function in redshift-space was computed
using the  following estimator (Hamilton, 1993):
 \begin{equation} 1 +
\xi(s)=
\frac{DD(s) RR(s)}{DR^2(s)} \end{equation} where DD is the number of
the galaxy-galaxy pairs, DR is the number of galaxy-random point pairs
and RR is the number of pairs in the random catalog with a separation
in the interval $s$ and $s + ds$. This estimator is less sensitive to
uncertainties of the mean density, which is a second order effect in
contrast to the previously defined  estimator (Davis \& Peebles 1983).
Since we only consider samples limited both in distance and in
absolute magnitude we adopt uniform weighting in the pair-counts
defined above.
From these pair counts in shells, we have also computed the so-called 
$J_{3}$ integral (Peebles 1980) 
\begin{equation} 
J_3(s) = \int_{0}^{s} u^2 \xi(u) du
\end{equation}  
which represents (up to a factor of $4\pi$) the mean number of excess
galaxies around each galaxy within a distance $s$. We use this
statistic in section 4 to derive the relative bias between different
luminosity classes.

Another way to characterize the clustering properties of galaxies consists
in calculating the probability $P(N,l)$ of finding N galaxies in spherical 
cells of radii $l$ placed at random. In this procedure, the spheres crossing 
the edges of the samples are rejected. The number of 
trials is chosen in order to have small statistical errors at scales
larger than $2 h^{-1}$ Mpc (see the discussion in Appendix B of 
Maurogordato, Schaeffer, da Costa, 1992). This leads in our case to $10^4$
to $10^6$ trials, from the shallower to the deeper sample. We have derived 
the variance of these counts in cells which can be expressed as
\begin{equation}
<(N- \bar{N})^2> = \bar{N} (1+N_c),
\end{equation}
where n is the mean density of the sample, $\bar{N} = nV$ is the mean number
of objects, and $ N_c =n^2V^2\sigma^2$ is the mean number of objects in excess
of random  inside a sphere of volume V. The variance can be related to the
volume-averaged correlation function
\begin{equation}
\sigma^2(s) \equiv \bar{\xi}(s) = \frac{1}{V^2} 
{\int_V {\xi (s) d^3 s_1 d^3 s_2}}
\end{equation}

In the power law approximation, $\xi(s)=(s_0/s)^\gamma$,
this can be expressed as 
\begin{equation}
J_3(s)/V= \displaystyle{\frac{1} {(1-\gamma/3)}} \xi(s)
\end{equation}
\begin{equation} 
\sigma^2 (s) =  
\displaystyle{\frac {2^{-\gamma}} {(1-\gamma/3)
(1-\gamma/4)(1-\gamma/6)} }
\xi(s)
\end{equation}
In this case $ J_3(s)/V $ and $\sigma^2(s)$  are related by a
multiplicative  factor depending weakly on $\gamma$ in the range of
interest  (0.8 for $\gamma=1$, 0.75 for $\gamma=2$). Below we will compare
the results obtained from both methods to test our conclusions.

\subsection{Volume and Luminosity limited Samples}

From the SSRS2 we have extracted nine volume and luminosity limited
subsamples with their absolute magnitude limit separated by $0.5^m$,
ranging in depth from $30$ to $168$ \h. The characteristics of these
samples are given in Table 1 where we list: in column (1) sample
identification; in column (2) the depth as defined in the previous
section; in column (3) the corresponding faint absolute magnitude
limit of the sample; in column (4) the number of galaxies in the
sample; in column (5) the mean density of galaxies.

\placetable{tbl-1}

For each of these samples we have computed $\xi(s)$, $J_3 (s)$ and
$\sigma^2(s)$, and derived power-law fits to $\xi(s)$ in the range of
separations where it is a good approximation (this range is not the same for
all the samples). The correlation parameters derived from these
fits are listed in columns (6) and (7) of Table 1 where we give the
correlation length $s_0$ and slope $\gamma$ for a two-parameter fit in
the separation range range given in column (9). For comparison we list
in column (8) the correlation length $s_0^{\prime}$ obtained assuming a
fixed slope $\gamma = 1.7$. In column (10) we give the value
of the variance in $8 h^{-1}$ Mpc spheres, $\sigma_8^2$ as derived
from the analysis of counts in cells. Finally, in column (11), we list
the ratio $L/E$ of late type over early type galaxies.

\placefigure{fig1}

Figure 1 shows $\xi(s)$, $J_3(s)$ and $\sigma^2(s)$ for the samples
given in Table 1. The error bars correspond to the rms value as
measured for 15 bootstrap re-samplings of each sample. In the case of the less
 dense samples,  errors on  the 
various statistics on small scales can be very large. Therefore, here and in 
the following graphs, error bars for these samples are shown as dotted and 
dashed lines in order to avoid over-crowding of the plots. While we do not find
any systematic variation of the slope with depth (table 1), the
correlation length varies significantly over the range of depths
considered especially for the deepest samples containing the brightest
galaxies.  The observed variation is consistent to that shown in Davis
\etal (1988) but now probing much larger depths for a given absolute
magnitude limit. The same trend for an apparent increase in the
clustering strength with depth is seen in the number of neighbors
$J_3(s)$ and in the variance computed from counts-in-cells. This
variation is quantified by $\sigma_8^2$ listed in table 1.

A key question is whether the observed variation reflects the
intrinsic nature of the galaxy distribution (\eg
Baryshev  et al. 1995 and references therein), simply sampling variations 
or genuine luminosity segregation,
since in deeper samples the fraction of intrinsically luminous
galaxies increases. The discussion is particularly problematic since
comparison of the amplitude of the correlation function as determined
from different volumes may be affected by sampling variations which
can arise because different structures are probed for different
volume-limited samples.  This could induce fluctuations in the mean
density of the sample or variations in  redshift distortions as
the sample probes different virialized systems and samples different
regions of large-scale peculiar velocity field. Still another effect
that could be introduced by redshift distortions would be if more
luminous galaxies were found preferentially within virialized systems,
as a product of mergers for instance. In this case, redshift
distortions could also affect the comparison of samples even within
the same volume.

Ideally, to prove the existence of luminosity segregation one should
compare the clustering properties of different luminosity classes
within the same volume and in real space.  Recently, this has become
partially possible thanks to the large number of galaxies in complete
dense surveys like the SSRS2. The large number of galaxies allows the
sample to be divided in different ways to discriminate between the
various effects discussed above. This can be achieved by comparing the
clustering properties of galaxies of different luminosity classes
within the same volume and the clustering of galaxies in the same
luminosity class but in different volumes.  Although dividing the
sample in this fashion reduces the number of galaxies in each
subsample, thus increasing the uncertainties of the various
statistics, it makes it possible to distinguish the contribution that
different effects may have on the \xis derived from volume-limited 
samples like those used in figure 1.

\placefigure{fig2}

In figure 2 we show \xis for three different volumes ($D=38h^{-1}$
Mpc, $D=60h^{-1}$ Mpc, $D=91h^{-1}$Mpc).  For each volume we have
compared the clustering properties of galaxies in contiguous magnitude
bins $1^m$ wide. Unfortunately, the sample is not large enough to
allow this comparison to be made over a wider range of luminosity
classes since we are constrained by the apparent magnitude limit of
the sample and by the number of galaxies in each class.  In our case
we are limited to only two adjacent magnitude bins given in Table 2.
The depth was chosen to correspond to the maximum distance at which
the faintest galaxies considered are visible. Dividing the sample 
in this way allows us to test for luminosity segregation in a way 
that is insensitive to sampling variations.

\placetable{tbl-2}

We note from figure 2 that faint galaxies do not exhibit any
significant relative bias. While there is some slight indication of
segregation for galaxies in the range $-20.5 \leq M \leq -19.5$
relative to galaxies in the magnitude interval $ -19.5 \le M \leq
-18.5$ (panel b), it is only in the largest volume considered that we
find a significant effect. This result implies  that galaxies
brighter than $-20.5$ are significantly more clustered than those in
the range $-20.5 < M < -19.5 $. Based on these results we infer that
the onset of this segregation occurs roughly at M = -19.5, close to
$L_*$.

\placefigure{fig3}

In order to examine the importance of sampling variations we show in
figure 3 measurements of \xis for galaxies in the same luminosity
class in three different volumes of increasing depth. The
characteristics of these subsamples are also summarized in Table 2.
From the plot we find that the clustering of faint galaxies seen only
in relatively nearby samples is not stable which reflects strong
fluctuations in the mean density.  This is clearly seen in panel (a)
where we show $\xi(s)$ for galaxies in the luminosity range $ -19.5
\leq M \leq -18.5$ in volumes ranging from 38 to 60 \h in depth as
indicated in the figure. For the faintest range considered the volumes
are relatively small and one finds some variation in the correlation
function at large separations with the amplitude of $\xi(s)$ computed in
the $60$ \h volume being significantly larger for $s > 6$ \h. This is
not surprising since the size of the samples are comparable to the
scale of inhomogeneities leading to important fluctuations in the mean
density. This is not the case for the brighter subsamples probing
volumes with depths $D >$ 60 \h . In fact, although one can see in
figure 1 (a) an increase in the amplitude of the correlation function
between, for instance, samples D38 and D60, no significant luminosity
segregation is detected when comparing the $\xi(s)$ for different
luminosity classes within the same volume.  Instead $\xi(s)$ is seen
to vary significantly with the depth of the volume considered.

By contrast, for galaxies brighter than $M=-19.5$, shown in panels
(b) and (c) there is no evidence for variations in the correlation
strength with the depth of the sample. This suggests that sampling
variations affect our results only for samples that probe small
volumes of space.  This conclusion is confirmed by the fact that the
ratio of $\xi(s)$ for galaxies in the D91 subsample and for galaxies
in the sample D138 is consistent with the ratio of $\xi(s)$ for the
same luminosity classes within the same volume.  This indicates that
luminosity segregation is responsible for the increase in the
amplitude of $\xi(s)$ between sample D91 and D138 seen in Figure 1.
We assume that the same is valid for galaxies brighter than $M=-21$,
although we cannot verify this claim because the number of such bright
galaxies rapidly decreases for smaller volumes. Nevertheless,

we have checked that the mean density within this sample, calculated in 
shells of
different depths, does not show any systematical variation. In particular, no
drop of the density with depth has been detected. Therefore the clustering 
properties
of the very bright galaxies previously shown are not  a spurious effect due 
to an hypothetical  incompletness.
The properties of the subsample D168 are
examined in detail in a forthcoming paper (Cappi \etal 1996).

Based on the previous discussion we conclude that the variation of the
amplitude of the correlation function observed in Figure 1 has two
contributions. The first due to sampling fluctuations which affects
primarily the clustering of faint galaxies probing small volumes. This
affects samples roughly to D74 or galaxies fainter than $M=-19$.  The
second effect is consistent with the interpretation that the
clustering of galaxies depends on luminosity.  In the discussion
below, we use the correlation function of volume-limited samples
including all galaxies brighter than a limiting value to derive the
relative bias of galaxies as a function of luminosity. We do that
because, although useful to investigate the onset of the segregation,
comparison of galaxies in magnitude bins are not suitable for
comparison with theoretical predictions.

\subsection {Redshift Distortion and Morphology Effects} As different
structures are probed by the samples used in figure 1, variations in
the
amplitude of the correlation could also be induced by redshift
distortions. On small scales, virialized systems tend to lower the
correlation amplitude, while on intermediate scales streaming motions
enhance the correlation amplitude, when measured in redshift space
(Kaiser 1987). However,  as da
Costa \etal (1995) have shown that redshift distortions are relatively weak 
within the
volume surveyed by the SSRS2, we do not expect a strong impact on the 
statistics.
Nevertheless, in order to quantify this effect, we have checked the behavior
of the correlation function in real space on the previously tested samples. 
We have verified that the trend with luminosity is
the same, and that the amplification factor between the various samples
is quantitatively the same in real space as in redshift space. This shows that
redshift space distortion are not responsible for the luminosity 
biasing evidenced by SSRS2 data.  

We note that our results cannot be due to a
morphology-luminosity relation since Marzke \etal (1994) and da Costa
\etal (1995) have shown that the luminosity functions for different
morphological types are similar, except for irregulars. Moreover, we
have verified that the morphological composition of our volume-limited
samples is independent of the luminosity class adopted (Table 1, column (11)).

\subsection{Comparison with previous results}

The above results give further support to conclusions reached by
Hamilton (1988), Davis \etal (1988) and Lachi\`eze-Rey, da Costa and
Maurogordato (1992) from their analyses of the CfA1 and SSRS redshift
surveys, with significantly better statistics and extending the range
of magnitudes over which the effect can be studied. First, both
sampling effects and stronger clustering of more luminous galaxies are
responsible for the variation of the correlation length with the depth
of the sample. Sampling effects are important for the smaller volumes
and are difficult to disentangle from luminosity effects below  $M \sim -19$.
However, luminosity segregation at the low-luminosity end, if present at all, 
should be small, as figure 2 shows no evidence for a variation of the
clustering amplitude of galaxies in the magnitude interval $-18.5 \leq
M \leq -17.5$ relative to those in the interval $-19.5
\leq M \leq -18.5$ within the volume of depth 30\h . Second,
for brighter galaxies, probing larger volumes, sampling effects are  not
important and for a given luminosity limit there is no significant
variation of the correlation amplitude with depth. On the other
hand, comparison of different luminosity classes within the same
volume shows robust evidence for a relative increase in the clustering
strength as a function of the luminosity. The relative amplitude is
the same as that deduced from considering galaxies in the magnitude
bins in the same volume.  Third, the values of the relative bias when 
comparing the brightest galaxies in the sample to galaxies
fainter that $L_*$ are  in good agreement with  the values obtained by
Hamilton (1988) from the CfA1.

Signatures of luminosity segregation have also been detected by Park
\etal (1994) from their power spectrum analysis of volume-limited
subsamples drawn from CfA2. In particular, the amplification factor of
$\sim 1.4$ between their subsamples at $101 h^{-1}$ Mpc, $M \le -19.7$
and $130 h^{-1}$ Mpc, $M \le -20.3$ is consistent with our findings,
However, their comparison of the power spectrum from bright/faint
galaxies in volume-limited samples is not easily compared to our
results as they use different magnitude bins.

Evidence for luminosity-dependence of clustering have also been
detected by Loveday \etal (1995) from a recent analysis of the correlation
function on the Stromlo$-$APM redshift survey. Their conclusions differ from
our findings as they show that while galaxies
brighter than $L_*$ ($-20 \le M \le -19$) are more clustered than
fainter ones ($-19 \le M \le -15$), there is no evidence for stronger
clustering of galaxies with $L > L_*$ ($-22 \le M \le -20$).
The reasons for these differences are at the present unclear.

\section{Bias and Comparison to Models}

 In order to quantify the dependence of the clustering
strength on luminosity we define the relative bias $b/b*$ of galaxies 
brighter than
$M$ towards galaxies brighter than $M_*$ ($\sim -19.5$) as a function of scale
$s$:

\begin{equation}
\displaystyle{\frac{b}{b_{*}}}(s)=\displaystyle{\sqrt{\frac{\xi(s)}{\xi_*(s)}}}
\end{equation} 
We have adopted the following notation for the various statistics
applied: no subscript refers to galaxies brighter than $M$, and $*$
subscript refers to galaxies brighter than $M_*$.

On those scales where the correlation function is well fitted by a
power law, $J_3(s)/V$ and $\sigma^2(s)$ are proportional to $\xi(s)$,
and therefore the ratio $b/b_*(s)$ can alternatively be expressed in
terms of $J_3(s)$ or of $\sigma^2(s)$ as

\begin{equation}
\displaystyle
{\frac{b}{b_{*}}}(s)=\displaystyle{\sqrt{\frac{J_3(s)}{J_{3}^*(s)}}}
\end{equation}
 
\begin{equation}
\displaystyle{\frac{b}{b_{*}}}(s)=\displaystyle{\sqrt{\frac{\sigma^2(s)}
{\sigma^{2}_*(s)}}}
\end{equation} 

It is more convenient to work with expression (10) and (11) rather than with 
expression (9) as $J_3(s)$ or $\sigma^2(s)$, which are integrated statistics, 
make possible the derivation of smoother ratios.

As mentioned above, while the use of magnitude bins is useful to
detect
and to determine the onset of luminosity segregation, for comparison
with theoretical predictions it is more convenient to use volume and
absolute magnitude limited samples like those of Table 1.
Results relative to $M>-19.0$ should be considered
with caution before further interpretation, as sampling effects have been 
shown to be important for the smaller samples.

\placefigure{fig4}

The ratio $b/b_{*}$ as a function of scale is presented in figure 4 as
derived from ratios of $J_3(s)$ (panel a) and $\sigma^2(s)$ (panel b). It
is worth noting that the ratio is roughly constant in the interval $2
< s < 13 h^{-1}$ Mpc for the subsamples considered, indicating that the 
relative
bias is independent of scale, which makes it possible to define a mean linear 
biasing factor, which varies from 0.8 for the faintest sample ($M \leq -19$)
to 2 for the brightest sample ($M \leq -21.0$).

The variation of the linear bias with luminosity is shown in figure 5,
where we show for comparison the same curve obtained by Hamilton
(1988) from the CfA1 survey. As can be seen there is a remarkable
agreement at the bright end between these estimates based on samples
probing different structures. It is important to
emphasize that the SSRS2 sample is not affected by uncertainties in
the Virgo infall which has been used as an argument against Hamilton's
interpretation of his results based on CfA1 data.

\placefigure{fig5}

 Luminosity biasing is expected to be present in the galaxy
distribution according to most scenarios of galaxy formation.  In the
standard scheme of biased galaxy formation (Bardeen et al. 1986),
galaxies are assumed to form from peaks above some global threshold in
the primordial smoothed density field . This implies a dependence of
the correlation function on the threshold which can be converted into
a dependence on luminosity once a luminosity or circular velocity-peak
relation is determined. Direct evidence that the amplitude of the
correlation function increases with the circular velocity of the halos
has been found both in numerical simulations of hierarchical
clustering (White et al 1987) and in an extension of the
Press-Schechter formalism (Mo and White 1995). Comparing these
predictions to our data requires the knowledge of the relation between
halos and galaxies. We have compared our results to these predictions
by expressing circular velocities in terms of luminosities by the
Tully-Fisher and Faber-Jackson relations (White, Tully and Davis 1988
and Valls-Gabaud, Alimi and Blanchard 1989).  We have normalized the
results for all the samples tested to the sample with galaxies
brighter than $L*$. As shown in figure 5, even in the case of low
normalization ($\sigma_8$=0.5), it is difficult to mimic the steep
increase of clustering present in the data at $L > L*$. Moreover, all
models exhibit a clear luminosity dependence also at low luminosities
which is difficult to test from the data as low-luminosity samples are
also nearby and have been shown to be sensitive to sampling
fluctuations. Still, as discussed before, the lack of luminosity
segregation inside single volume-limited nearby samples as shown in
figure 3 tends to indicate that a {\it strong} segregation at the
low-luminosity end is quite unlikely.  Note that this comparison is
only suggestive as a one-to-one correspondence has been assigned to
galaxy-halo pairs.  More elaborated modeling of the properties of
galaxies and halos (Kauffmann, 1995) will allow a more precise
comparison of luminosity biasing to the prediction of models for {\it
galaxies}.

An alternative view on the origin of biasing can be obtained from a
description of the density field behavior at scales where it is
strongly non-linear (Schaeffer 1984, 1985, Balian \& Schaeffer 1989).
This approach yields an analytic description of the matter
fluctuations, based on an assumption about the scaling of the
many-body matter correlation functions inspired by dynamical
considerations (Peebles 1980, Balian \& Schaeffer 1989). These models
imply a bias in the distribution of high density spots. However, this
bias is not acquired at the birth of the over-densities as a result of the
initial Gaussian conditions, and then frozen. It is the result
(Schaeffer 1985, 1987) of the non-linear evolution, and is determined
once the density fluctuations are known. A specific calculation
(Bernardeau \& Schaeffer 1992) shows that for all models with the
above scaling, the bias depends only on the internal properties of the
objects whose correlations are considered.  More specifically it
increases as a function of a unique parameter, closely related to the
gravitational potential, and therefore increases with luminosity. This
analytical model gives explicitly the asymptotic behavior for the
brightest galaxies: $b/b^*$ is nearly proportional to $L/L^*$. It is
very successful in reproducing the steep increase of the correlation
function for bright galaxies $L > L*$ but it predicts a strong
luminosity segregation at faint luminosities which is very difficult
to reconcile with the data (figure 5).  Larger samples of galaxies
going to fainter magnitudes are therefore necessary to test the models
at the low-luminosity end.

\section{Summary}

Using the recently completely SSRS2 redshift survey we have
investigated the dependence of the clustering properties of galaxies
on luminosity. The SSRS2 sample is particularly suitable for this
study because of the large number of galaxies, the paucity of rich
clusters within the volume and the small amplitude of redshift
distortions induced by peculiar motions both at small and large
scales. Analysis of volume-limited samples using both the two-point
correlation function and the variance of counts in cells indicate that
galaxy clustering depends on luminosity. The effect is small for faint
galaxies but increases dramatically for galaxies brighter than $L_*$.
The correlation length varies from about 5 \h, for the faintest
galaxies up to 8 \h for the galaxies brighter than $M=-20.5$. For the
brightest galaxies in the sample with $M \leq -21$ the correlation
length reaches values as high as $\sim 16$ \h. These galaxies are not
preferentially found in rich clusters and the study of their
properties  will be the subject of a separate paper. The relative
bias is scale independent up to $\sim 13$ \h and the linear bias
between the brightest and $L_*$ galaxies is in the range 1.5-2.

 A preliminary comparison with presently available models of
galaxy formation shows serious difficulties in modeling simultaneously
the faint and bright ends of the bias variation with luminosity
evidenced in the SSRS2. However, more data is necessary to allow a
better determination of the behavior of luminosity biasing at faint
luminosities which is strongly affected by sampling variations for
nearby catalogs, although it is constrained to be small by our
analysis.  On the theoretical side, a better understanding galaxy formation
and of the galaxy-dark-halo connection is required in order to use
these results as useful constraints to galaxy formation models.

\bigskip
\acknowledgments

We would like to thank all the SSRS2 collaborators for allowing us to
use the data in advance of its publication.  We are grateful to George
Efstathiou, Guinevere Kauffmann,  Amos Yahil and Simon White for
fruitful discussions. LNdC would like to thank the hospitality of the
IAP and of the DAEC where most of the work was carried out.
We would like to thank the referee , Jo Ann Eder, for her constructive
comments.

\newpage

\figcaption [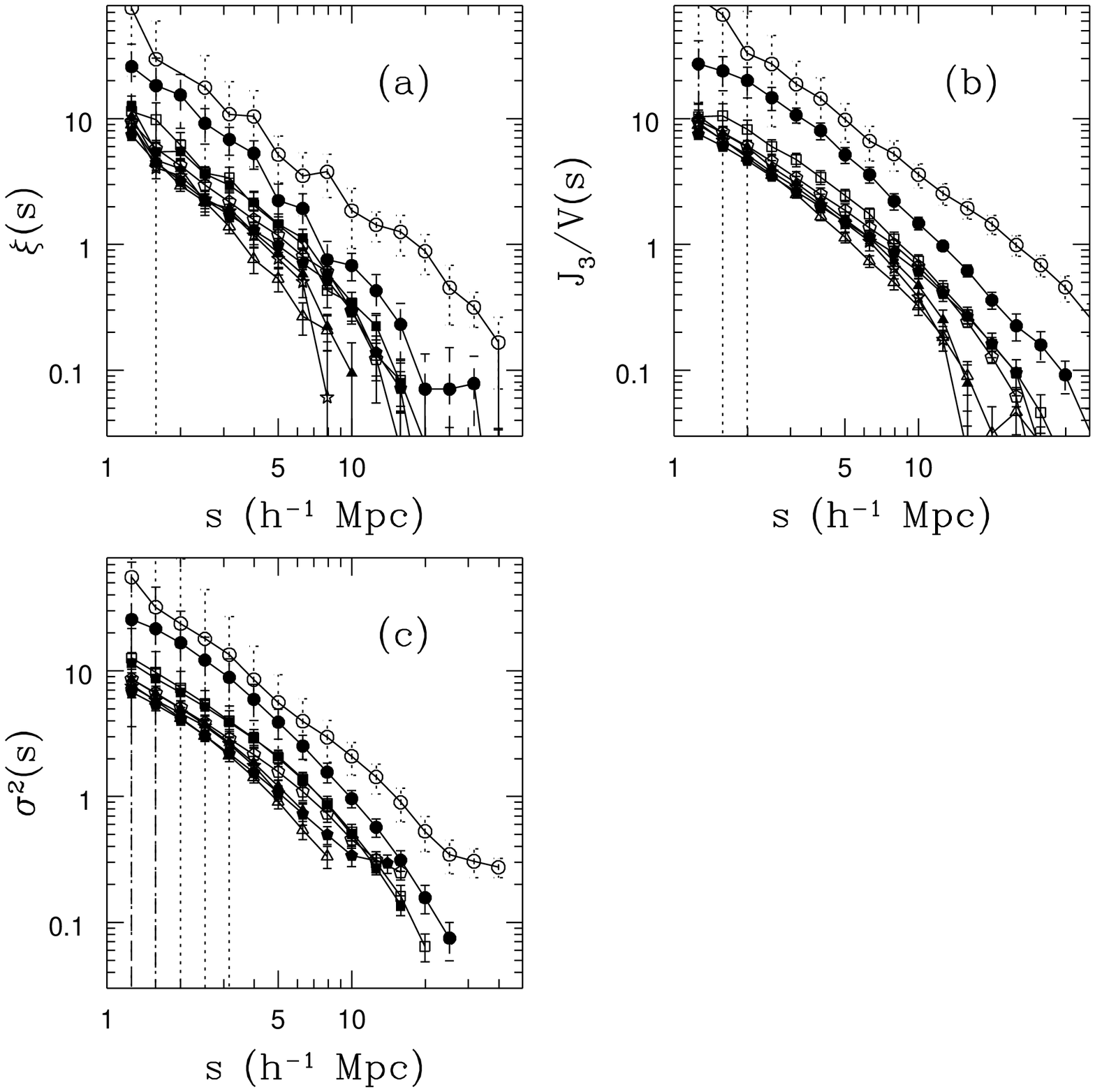]{The two-point correlation function in redshift
space $\xi(s)$ (panel a), the $J_3(s)$ integral (panel b) and the
variance $\sigma^2(s)$ (panel c) calculated from counts in cells
for SSRS2 volume-limited subsamples identified as follows : D30
(stars); D38 (full triangles); D48 (opened triangles): D60 (full
pentagons): D74 (opened pentagons); D91 (full squares); D112 (open
squares); D138 (full circles); D168 (open circles).\label{fig1}}

\figcaption[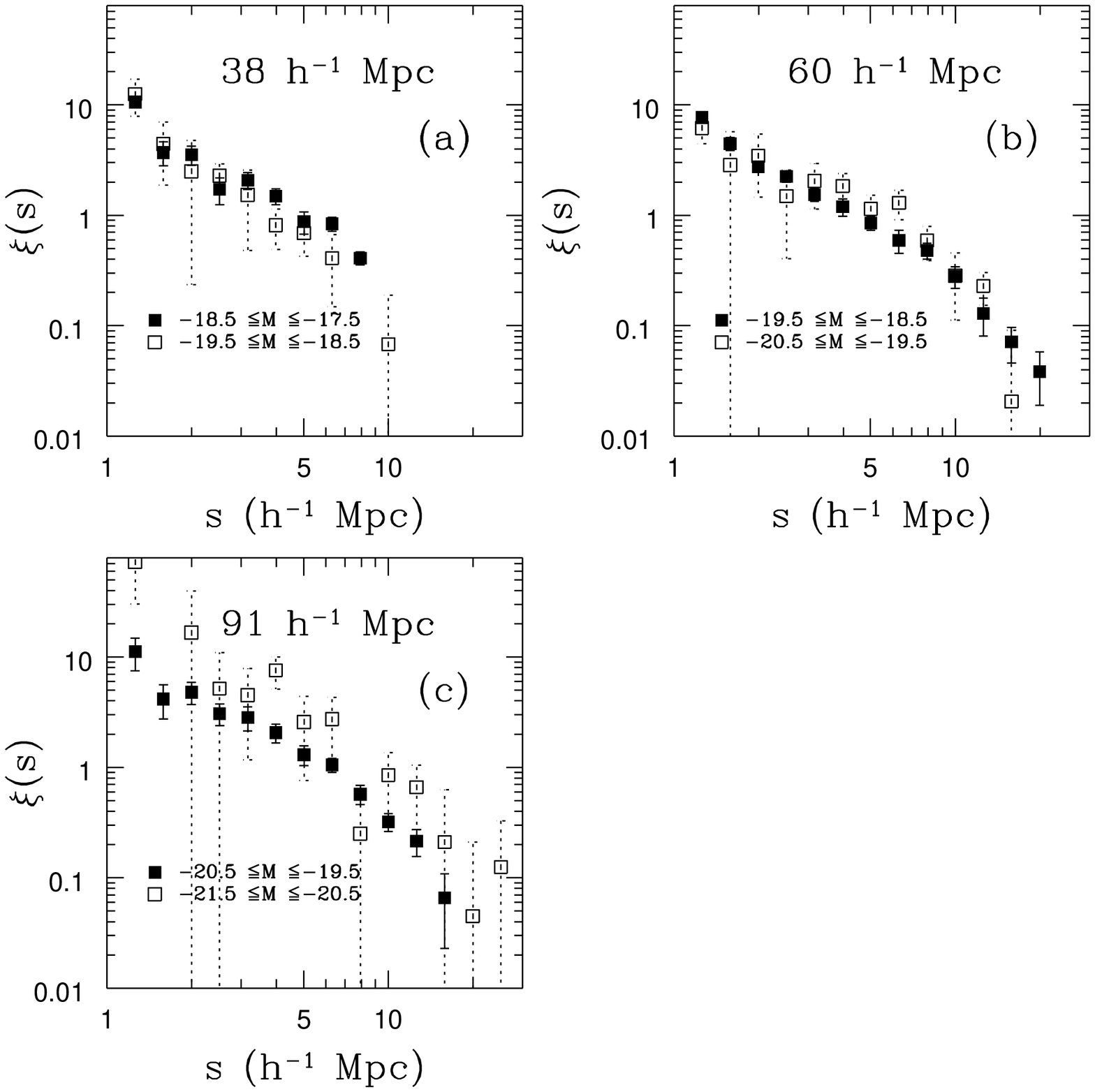]{The variation of $\xi(s)$ with luminosity is tested
within subsamples at a given depth. In each panel, we present
$\xi(s)$ for two luminosity classes of one magnitude range within the
same volume.  In panel a), galaxies with $ -18.5 < M < -17.5$ and
$-19.5 < M < -18.5$ are compared within a volume of $38h^{-1}Mpc$ depth.
In panel b), galaxies with $ -19.5 < M < -18.5$ and $-20.5 < M <
-19.5$ are compared within a volume of $60h^{-1}Mpc$ depth.  In panel c),
galaxies with $ -20.5 < M < -19.5$ and $-21.5 < M < -20.5$ are
compared within a volume of $91h^{-1}Mpc$ depth.}

\figcaption[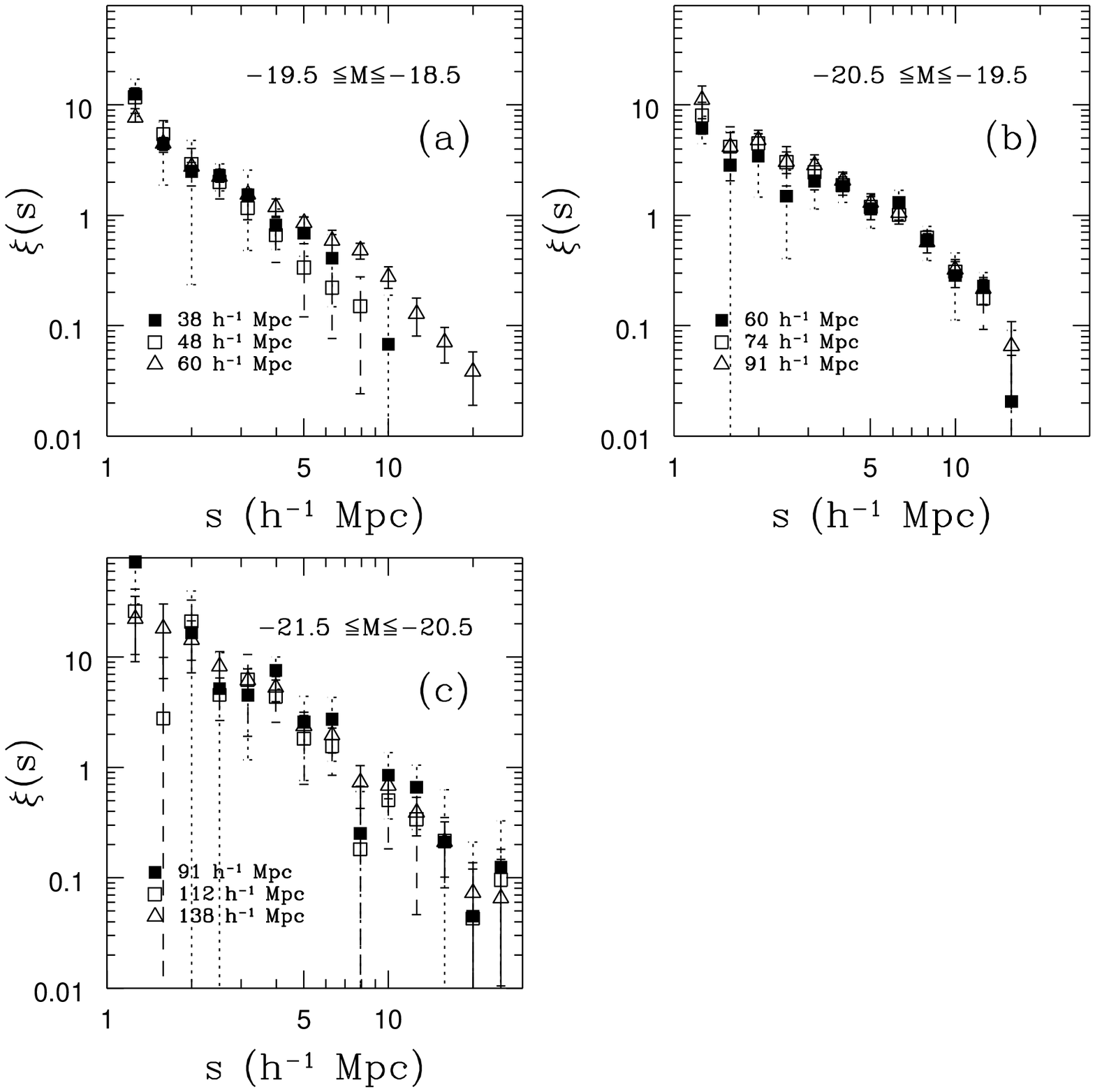]{The redshift space correlation function
$\xi(s)$ is displayed in various panels, each corresponding to a
given luminosity class: $-19.5< M < -18.5$ in panel a), $-20.5< M < -19.5.$ 
in panel
b), $-21.5< M < - 20.5$ in panel c). Within each panel, $\xi(s)$ is presented
for three different volumes. Symbols vary from opened triangles
to opened squares and full squares with increasing size of the
sample. }

\figcaption[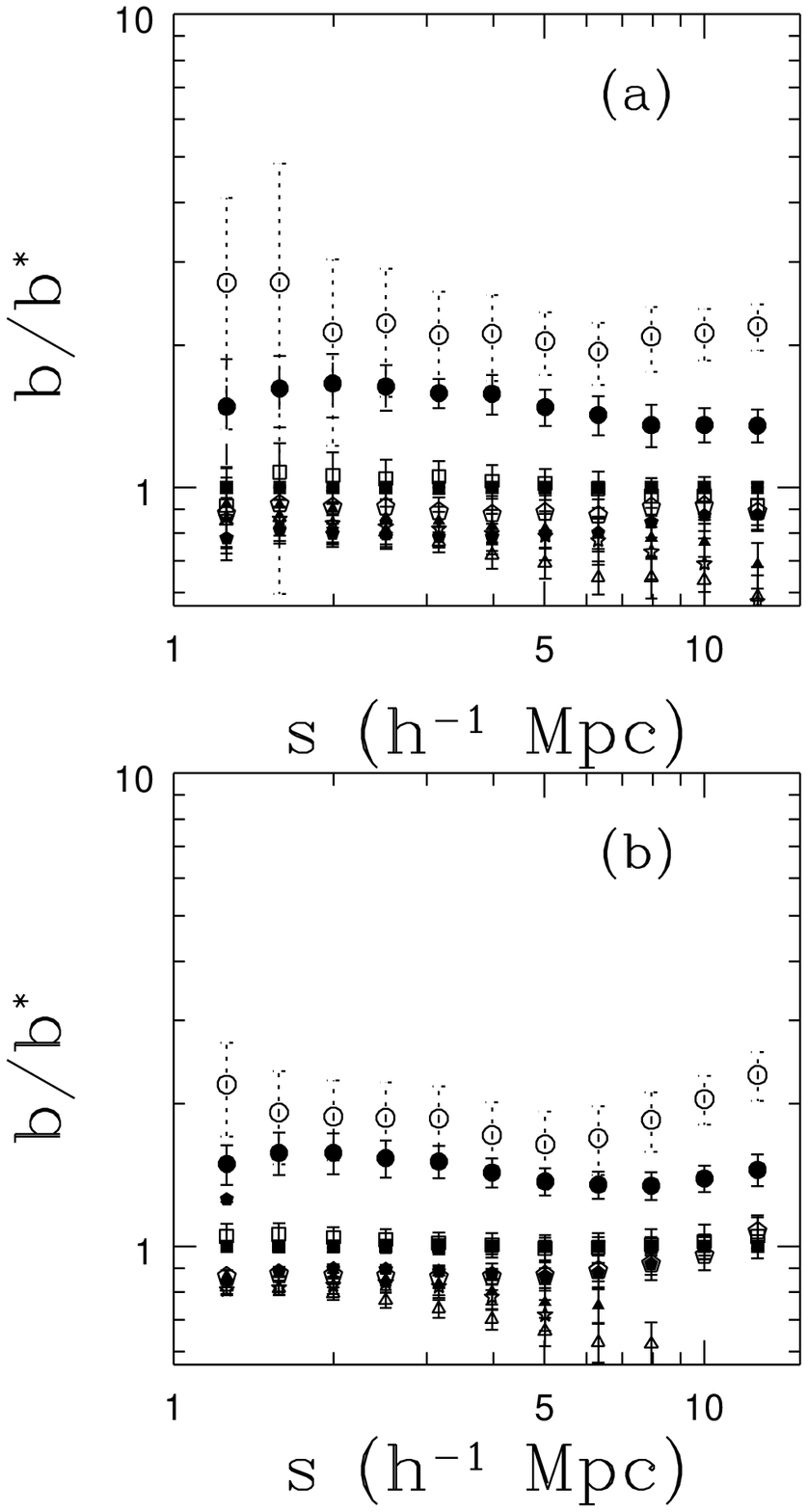]{The biasing factor as a function of separation, 
for subsamples D74 to D168. It is normalized to the value $b^{*}$
calculated from sample D91 corresponding to galaxies brighter than
$L^{*}$. In panel a), the biasing factor is derived from $J_3$, 
and in panel b) from $\sigma^2$. }

\figcaption[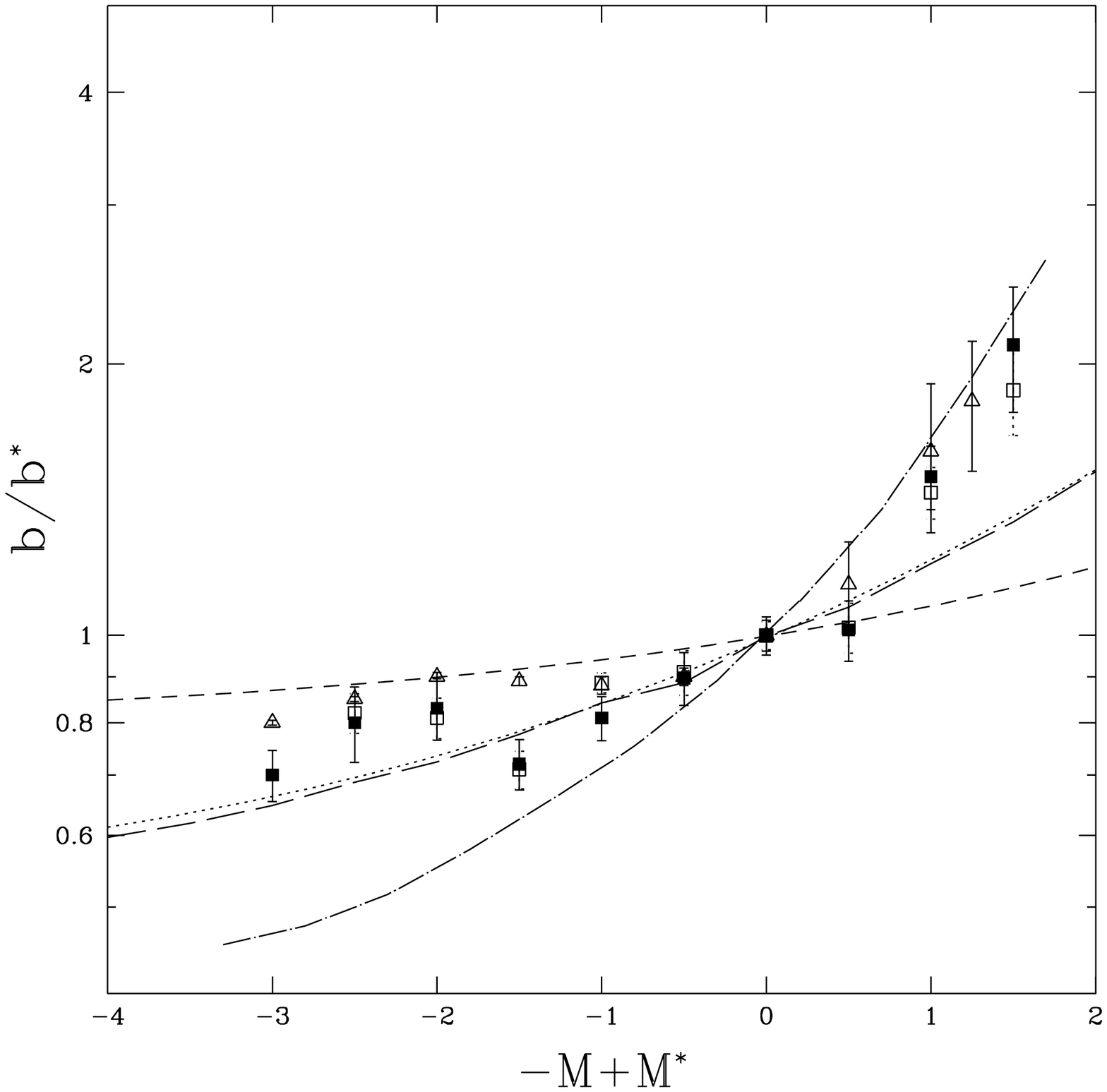]{Evolution of $b/b^{*}$ (averaged
 between $2h^{-1}Mpc$ and $13h^{-1}Mpc$) with the limiting absolute
 magnitude of the sample . Our results from SSRS2 are reported as
full squares when derived from $J_3$, and as open squares when derived 
from $\sigma^2$. CfA results  of Hamilton (1988) are shown  as opened 
triangles. Results from CDM simulations of  White \etal (1987) as given by 
Valls-Gabaud et al. (1989) are plotted as a  long-dashed line.
On this graph are over-plotted analytical predictions by  
Mo and White (1995), with low ($\sigma_8 = 0.5 $, dotted  line) 
and high normalizations ($\sigma_8 = 1$, short-dashedline) 
and by Bernardeau \& Schaeffer (1992) (dashed-dotted line).}

\newpage
\begin{center}
\begin{deluxetable}{ccccccccccc}
\scriptsize
\tablecaption{Subsamples of the SSRS2. \label{tbl-1}}
\tablewidth{0pt}
\tablehead{
\colhead{Sample} & \colhead{D}   & \colhead{$M_{lim}$}   & \colhead{ N$_{g}$} & 
\colhead{$\bar{n}$ }  & \colhead{$s_{0}$} & \colhead{$\gamma$} 
& \colhead{$s_0\prime$} &\colhead{ Interval } &\colhead{$\sigma_8^2$ } 
&\colhead{$L/E$}
\nl
 \colhead{}&\colhead{($h^{-1}$Mpc)} & \colhead{} &\colhead{}    
& \colhead{($10^{-3}h^3$Mpc$^{-3}$)}&\colhead{($h^{-1}$Mpc)}
&\colhead{}& \colhead{($h^{-1}$Mpc)}&\colhead{($h^{-1}$Mpc)} 
& \colhead{} &\colhead{}
}

\startdata
D30 &30.69 &-17.0 &301 &27.2 &4.3$\pm$0.5 &1.7$\pm$0.2 & 4.3$\pm$0.4 
&2.0-5.2 &- &2.3
\nl
D38 &38.35 &-17.5 &329 &15.5 &4.5$\pm$0.4 &1.6$\pm$0.2 & 4.4$\pm$0.3 
&2.0-5.2 &- &2.2
\nl
D48 &47.83 &-18.0 &487 &11.8 &3.5$\pm$0.2 &2.1$\pm$0.2 &3.5$\pm$0.2 
&2.0-5.0 &0.3 &2.2
\nl 
D60 &59.52 &-18.5 &774 &9.7 &4.8$\pm$0.5 &1.3$\pm$0.1 & 4.6$\pm$0.3 
&2.0-8.0 &0.5 &2.0
\nl 
D74 &73.86 &-19.0 &770 &5.1 &5.5$\pm$0.4 &1.4$\pm$0.1 &5.2$\pm$0.4 
&2.0-8.0 &0.7 &1.9
\nl
D91 &91.37 &-19.5 &755 &2.6 &5.8$\pm$0.6 &1.7$\pm$0.2 &5.8$\pm$0.4 
&2.0-12.6 &0.9 &2.1
\nl
D112 &112.60 &-20.0 &593 &1.1 &5.5$\pm$0.6 &2.0$\pm$0.3 &5.5$\pm$0.4 
&2.0-12.6&0.9 &2.2
\nl
D138 &138.12 &-20.5 &268 &0.3 &8.0$\pm$0.4 &2.0$\pm$0.2 &8.5$\pm$0.5 
&2.0-15.8 &1.5 &2.5
\nl
D168 &168.52 &-21.0 &133 &0.1 &15.8$\pm$2.9 &1.5$\pm$0.3 &15.2$\pm$2.0 
&2.5-25.0 &3.0 &1.7
\enddata
\end{deluxetable}
\end{center}

\newpage

\begin{center}
\begin{deluxetable}{ccc}
\scriptsize
\tablecaption{Characteristics of subsamples in magnitude bins.\label{tbl-2}}
\tablewidth{0pt}
\tablehead{
\colhead{Sample} & \colhead{D}  & \colhead{ N$_{g}$} 
\nl
 \colhead{}&\colhead{($h^{-1}$Mpc)} & \colhead{} 
}

\startdata
-18.5/-17.5 & 38.35 & 187  
\nl
-19.5/-18.5 & 38.35 & 105  
\nl
-19.5/-18.5 & 47.83 & 227  
\nl
-19.5/-18.5 & 59.52 & 563  
\nl
-20.5/-19.5 & 59.52 & 193  
\nl
-20.5/-19.5 & 73.86 & 329 
\nl
-20.5/-19.5 & 91.37 & 687  
\nl
-21.5/-20.5 & 91.37 &  67  
\nl
-21.5/-20.5 & 112.60 & 155 
\nl
-21.5/-20.5 & 138.12 & 268 
\enddata
\end{deluxetable}
\end{center}

\newpage

%%%%%%%%%      %%%%%%%
\begin{figure} [h]
\plotone{fig1.ps}
\figurenum{1}
   \caption{\label {fig1}}
\end{figure}
%%%%%%%%%%%%%%%%%%%%%%%%%%%%%%%

\newpage

%%%%%%%%%      %%%%%%%
\begin{figure} [h]
\plotone{fig2.ps}
\figurenum{2}
   \caption{\label {fig2}}
\end{figure}
%%%%%%%%%%%%%%%%%%%%%%%%%%%%%%%

\newpage

%%%%%%%%%      %%%%%%%
\begin{figure} [h]
\plotone{fig3.ps}
\figurenum{3}
   \caption{\label {fig3}}
\end{figure}
%%%%%%%%%%%%%%%%%%%%%%%%%%%%%%%

\newpage

%%%%%%%%%      %%%%%%%
\begin{figure} [h]
\plotone{fig4.ps}
\figurenum{4}
   \caption{\label {fig4}}
\end{figure}
%%%%%%%%%%%%%%%%%%%%%%%%%%%%%%%

\newpage

%%%%%%%%%      %%%%%%%
\begin{figure} [h]
\plotone{fig5.ps}
\figurenum{5}
   \caption{\label {fig5}}
\end{figure}
%%%%%%%%%%%%%%%%%%%%%%%%%%%%%%%

\newpage


\newpage

\begin{thebibliography}{99}

\bibitem{} Alimi, J-M., Valls--Gabaud, D. \& Blanchard, A., 1988, 
A\&A, 206, L11

\bibitem{} Bardeen, J.M., Bond, J.R., Kaiser, N. \& Szalay, A.S., 1986, 
ApJ, 304, 15

\bibitem{} Balian, R. \& Schaeffer, R., 1989, A\&A, 220, 1

\bibitem{} Baryshev, Y.V., Sylos Labini, F., Montuori, M., \& Pietronero, L., 
1994, Vistas in Astronomy, 38, 419

\bibitem{} Bernardeau, F., \& Schaeffer, R., 1992, A\&A, 255, 1

\bibitem{} Bingelli, B., Tarenghi, M., \& Sandage, A., 1990, A\&A, 228, 42 

\bibitem{} B\"orner, G., Mo, H.J. \& Zhou, 1989, A\&A, 221, 191

\bibitem{} B\"orner, G., Deng, Z.G., \& Xia, X.Y., 1989, A\&A, 209, 1  

\bibitem{} B\"orner, G., \& Mo, H.J., 1990, A\&A, 227, 324

\bibitem{} Cappi, A., da Costa, N.L., Benoist, C. \& Maurogordato, S., 
in preparation

\bibitem{} da Costa, N.L., Pellegrini, P.S., Sargent, W.L.W., Tonry, J., 
Davis, M., Meiskin, A., Latham, D.W., Menzies, J.W. \& Coulson, J.A., 
1988, ApJ, 327, 544 

\bibitem{} da Costa, N.L., Geller, M.J.,Pellegrini, P.S., Latham,
D.W., Fairall, A.P., Marzke, R.O., Willmer, C.N.A., Huchra, J.P.,
Calderon, J.H., Ramella, M. \& Kurtz, M.J., 1994, ApJ Letters, 424, L1

\bibitem{} da Costa, L.N., Vogeley, M., Geller, M.J., Huchra, J., \&
Park, C., 1994, ApJ, 437, L1

\bibitem{} Davis, M., \& Geller, M.J., 1976, ApJ, 208, 13 

\bibitem{} Davis, M., \& Peebles, P.J.E., 1983, ApJ, 267, 465 

\bibitem{} Davis, M., Meiksin, A., Strauss, M.A., da Costa, N.L.,
 \& Yahil, A., 1988, ApJ, 333, L9 

\bibitem{} Dominguez-Tenreiro, R., \& Martinez, V.J., 1989, ApJ, 339, L9

\bibitem{} Dominguez-Tenreiro, R., Campos, A., G\'{o}mez-Flechoso, M.A., \& 
Yepes,G., 1994, ApJ, 424, L73  

\bibitem{} Eder, J.A., Schombert, J.M., Dekel, A., \& Oemler, A., 1989, ApJ, 
340, 29 

\bibitem{} Efstathiou, G., Ellis, R.S.\& Peterson, B.A., 1988, MNRAS,
232, 43

\bibitem{} Hamilton, A.J.S., 1988, ApJ, 331, L59

\bibitem{} Hamilton, A.J.S., 1993, ApJ, 417, 19

\bibitem{} Hasegawa, T., \& Umemura, M., 1993, MNRAS, 263, 191 

\bibitem{} Haynes, M.P., \& Giovanelli, R., 1989, in Large Scale Motion of the
Universe, ed. Rubin \& Coyne (Princeton U.P.), 31

\bibitem{} Huchra,J., Davis, M., Latham, D., \& Jonry, J., 1983, ApJS,
52, 89 

\bibitem{} Iovino, A., Giovanelli, R., Haynes, M.,Chicarini, G., \& Guzzo, L.,
 1993, MNRAS, 265, 21

\bibitem{} Kaiser, N., 1987, MNRAS, 227,1

\bibitem{} Kauffmann , G., 1995, private communication

\bibitem{}  Lachi\`{e}ze-Rey, M., da Costa, N.L., \& Maurogordato, S., 
1992, ApJ, 399, 10 

\bibitem{} Loveday, J., Maddox, S.J., Efstathiou, G., \& Peterson,
B.A., 1995, ApJ, 442, 457

\bibitem{} Marzke, R.O., Geller, M.J., da Costa, L.N., \& Huchra,
J.P., 1994, CFA preprint

\bibitem{} Maurogordato, S.,\& Lachi\`{e}ze-Rey, M., 1991, ApJ, 369, 30 

\bibitem{} Maurogordato, S., Schaeffer, R., \& da Costa. L.N., 1992, 
ApJ, 390, 17

\bibitem{} Mo, H.J., Einasto, M., Xia, X.Y., \& Deng, Z.G., 1992, 
MNRAS, 255, 382

\bibitem{} Mo, H.J., \& White, S.D.M., 1995, MNRAS, submitted

\bibitem{} Park, C., Vogeley, M.S., Geller, M.J., \& Huchra, J.P.,
1994, ApJ, 431, 569

\bibitem{} Peebles, P.J.E., 1980, The Large Scale Structure of the Universe 
(Princeton: Princeton Univ. Press)

\bibitem{} Phillipps, S., \& Shanks, T., 1987, MNRAS, 229, 621 

\bibitem{} Santiago, B.X., \& da Costa, N.L., 1990, ApJ, 362, 386

\bibitem{} Schaeffer, R., 1984, A\&A, 134, L15

\bibitem{} Schaeffer, R., 1985, A\&A, 144, L1

\bibitem{} Schaeffer, R., 1987, A\&A, 180, L5

\bibitem{} Thuan, T.X., Gott III, J.R., \& Schneider, S.E., 1987, ApJ, 315, L93

\bibitem{} Thuan, T.X., Alimi, J.M.,Gott, J.R., \& Schneider, S.E., 1991, ApJ, 
370, 25 

\bibitem{} Valls-Gabaud, D., Alimi, J.M., \& Blanchard, A., 1989, 
Nature, 341, 215 

\bibitem{} White, S.D.M., Davis, M., Efstathiou, G., \& Frenk, C.S.,
1987, 
Nature, 330, 351

\bibitem{} White, S.D.M., Tully, R.B., \& Davis, M., 1988, ApJ, 333, L45

\bibitem{} Yahil, A., Tammann, G.A., \& Sandage A., 1977, ApJ, 217,
903 

\end{thebibliography}
\end{document}